\documentclass[11pt]{article}
\usepackage{a4wide}                      
\usepackage{epsf}                        
\usepackage{latexsym}                    
\usepackage{amsfonts}                    
\usepackage{amssymb}                     
\usepackage{epsfig}			 
\usepackage{amsmath}                     

\def\be{\begin{equation}}
\def\ee{\end{equation}}
\def\bea{\begin{eqnarray}}
\def\eea{\end{eqnarray}}
\def\nn{\nonumber \\ [.2cm]}

\def\hsp#1{\hspace{#1}}

\def\part{\partial}

\def\Subst{\mbox{Subst}}  

\def\ud{\underline{d}}
\def\upartial{\underline{\partial}}

\def\makeatletter{\catcode`\@=11}
\makeatletter
\def\mathbox#1{\hbox{$\m@th#1$}}%
\def\math@ccstyles#1#2#3#4#5#6#7{{\leavevmode
      \setbox0\mathbox{#6#7}%
      \setbox2\mathbox{#4#5}%
      \dimen@ #3%
      \baselineskip\z@\lineskiplimit#1\lineskip\z@
      \vbox{\ialign{##\crcr
             \hfil \kern #2\box2 \hfil\crcr
             \noalign{\kern\dimen@}%
             \hfil\box0\hfil\crcr}}}}
\def\mathaccstyles{\math@ccstyles\maxdimen}
\def\maththroughstyles{\math@ccstyles{-\maxdimen}}
\def\unity%
 {\maththroughstyles{.45\ht0}\z@\displaystyle {\mathchar"006C}\displaystyle 1}
\begin{document}

\rightline{UG-FT-220-07}
\rightline{CAFPE-90-07}
\rightline{KUL--TF/07-26}
\rightline{December 2007}
\vspace{1truecm}

\centerline{\LARGE \bf Some thoughts about matrix coordinate transformations }
\vspace{1.3truecm}

\centerline{
     { \bf J. Adam${}^{a,}$}\footnote{E--mail address:
                                   {\tt joke.adam@fys.kuleuven.be}},
     { \bf B. Janssen${}^{b,}$}\footnote{E--mail address:
                                   {\tt bjanssen@ugr.es}},
     { \bf W. Troost${}^{a,}$}\footnote{E--mail address:
                                   {\tt walter.troost@fys.kuleuven.be} }
     {\bf and}
     { \bf W. Van Herck${}^{a,}$}\footnote{E--mail address:
                                   {\tt walter.vanherck@fys.kuleuven.be}}
                                                             }

\vspace{.4cm}
\centerline{{\it ${}^a$ Instituut voor Theoretische Fysica, K.U. Leuven}}
\centerline{{\it Celestijnenlaan 200D,  B-3001 Leuven, Belgium}}

\vspace{.4cm}
\centerline{{\it ${}^b$Departamento de F{\'\i}sica,  Universidad de
Granada,}}
\centerline{{\it Facultad de Ciencias, Campus de Fuentenueva, 18071 Granada, Spain}}

\vspace{2truecm}

\centerline{\bf ABSTRACT}
\vspace{.5truecm}
\noindent
Matrix coordinate transformations are defined as substitution operators without requiring an 
ordering prescription or an inclusion function from the Abelian coordinate transformations. We 
construct transforming objects mimicking most of the properties of tensors. We point out some 
problems with the matrix generalization of contravariant vectors. We suggest to use the 
substitution operators to search for an inclusion function.

\newpage

\section{Introduction}

Much of the non-perturbative knowledge about string theory is the result of studying extended 
objects like D-branes. In the case of a single D-brane, one can construct an effective action 
on the worldvolume of the D-brane, incorporating the low-energy description of open strings 
ending on this D--brane \cite{L}. In particular, the D-brane carries a $U(1)$ vector field, of 
which the degrees of freedom correspond to the massless vibration modes of the open strings 
tangential to the brane. On the other hand, the open string modes normal to the brane give rise 
to a set of scalars in the effective theory. These so-called embedding scalars are then 
interpreted as coordinates describing the embedding of the D-brane in the target spacetime. 

This picture changes dramatically in the presence of multiple coinciding D-branes. In the case of $N$ parallel D-branes at short distance (compared to the string length $\sqrt{\alpha'}$), extra 
bosonic degrees of freedom enhance the $N$ $U(1)$ symmetry groups of each object to a single 
$U(N)$ group \cite{W}. In this case, the $N$ brane components behave collectively and are 
rather described as a single object, which we will call the multiple brane. This $U(N)$ symmetry 
is reflected in the effective action of the multiple brane. In particular, the multiple brane 
carries a $U(N)$ Yang-Mills vector field with the same role as the $U(1)$ Born-Infeld vector of 
the single branes and instead of a set of real scalar fields the multiple brane carries a set of 
Hermitian matrices $X^\mu$ transforming in the adjoint of $U(N)$. Based on similarity with the 
single brane, the $n$-th eigenvalue of the $U(N)$ matrix $X^\mu$ is interpreted as the position 
of the $n$-th constituent brane in the transverse direction $x^\mu$. Such matrix coordinates open 
up the possibility of fuzzy brane configurations (see for example \cite{Myers}-\cite{CMT2}), 
but also quite some challenges.

In this article, we will focus on one of these challenges, which consists of coordinate 
transformations. Indeed, as the multiple D-brane is embedded in ordinary, relativistic 
spacetime, a coordinate transformation in the target space should leave the effective action 
invariant. The question of invariance brings up another, more basic question: how does a 
coordinate transformation affect a matrix coordinate? Most of the articles concerning matrix 
coordinate transformations, e.g. \cite{DS}-\cite{BKLvR}, search for a homomorphism between 
ordinary, Abelian coordinate transformations and (a subset of) matrix coordinate transformations. 
De Boer and Schalm \cite{DS} have pointed out that such a homomorphism requires a dependency on 
the background metric. So, one strategy would be to follow this lead and search for such an 
inclusion function, allowing a background metric dependence. Here, however, we will take a different route: instead we will forget about the Abelian coordinate transformations and try to construct 
an algebra of matrix coordinate transformations on its own. As we do not require a homomorphism 
between the algebra of Abelian coordinate transformations and our matrix coordinate transformations, we will not use any background metric. Actually, we will make no references to an Abelian 
background whatsoever.

So, our goal is to develop a group of matrix coordinate transformations, henceforth abbreviated 
by MCT. While we do not require a homomorphism from the Abelian coordinate transformations to 
the MCTs, we will use a projection from the MCTs unto the Abelian coordinate transformations. 
Such a projection is quite natural, as Abelian coordinates can be seen as ($1\times 1$)-matrices. 
We will require that this projection is a homomorphism from the MCT algebra unto the algebra of 
Abelian general coordinate transformations. In section \ref{sect_function}, we will define matrix 
functions to use as MCTs or as objects transforming under these MCTs. We define a substitution 
operator which serves as an infinitesimal MCT in section \ref{sect_scalar}.

Once we have defined consistent matrix coordinate transformations, we will look at their 
representations. In the Abelian case, we have a collection of invariant and covariant objects 
such as scalars, vectors and tensors. We want to create tensor-like objects transforming under 
the MCT representations and exhibiting most of the properties of the Abelian tensors. We will 
define a contravariant vector in section \ref{sect_contra}, thereby proving that the parameter 
of an infinitesimal MCT is indeed a contravariant vector. In the same section we will already 
point out some difficulties regarding scalar multiplication. In section \ref{sect_co}, covariant 
vectors and tensors are defined. We show how to get a structure similar to that of the 
antisymmetric differential forms using a differential operator not unlike the substitution 
operator defined in section \ref{sect_scalar}. We will look at contractions in section 
\ref{sect_contra_2} and draw attention to some problems that occur when defining dual spaces.

Lastly, we will discuss the MCTs and their transforming objects in section \ref{sect_discussion}. 
While the problems at the level of matrix vectors seem quite serious, we believe that the 
substitution operator and the matrix objects may help in the search for the inclusion function of 
De Boer and Schalm.

\section{The matrix function}\label{sect_function}

In the Abelian case, both the tensor fields and the coordinate transformations consist of (real 
or possibly complex) functions of the coordinates. In order to construct MCTs and transforming 
objects, we need to define functions of the matrix coordinates. There are many ways of defining 
these, one of them being for example a non-Abelian Taylor expansion \cite{GM}:
\be
F(X) = \sum_{k=0}^{\infty}\, \frac{1}{k!}\,\Bigl( \partial_{\lambda_1}...\partial_{\lambda_k}f \Bigr)
\vert_{x=0}\ X^{\lambda_1}...\,X^{\lambda_k}\,. 
\label{Taylorexp}
\ee
The coefficients in this expansion are derivatives of an Abelian function and therefore imply
a totally symmetric ordering of the $X^\mu$. However, this ordering prescription, and indeed 
any ordering prescription, does not fit with matrix coordinate transformations, as the 
composition $G = F_1 \circ F_2$ will not obey the same prescription as $F_1$ and $F_2$, as 
pointed out in  \cite{DS}. 

Instead of choosing a particular ordering prescription, as in (\ref{Taylorexp}), we will allow 
any ordering. A matrix function $F(X)$ is then defined by its expansion
\be
F(X) = \sum_{k=0}^{\infty}\, \frac{1}{k!}\, f_{\lambda_1...\lambda_k}\ X^{\lambda_1}...\,X^{\lambda_k}\,, 
\label{m_function}
\ee
where the coefficients $f_{\lambda_1...\lambda_k}$ are complex numbers. In contrast to the Abelian 
case, the coefficients need not to be symmetric for the exchange of indices. The product used 
in the definition of the matrix function~(\ref{m_function}) is the ordinary matrix product. 

Note that the image of such a function is not always a Hermitian matrix, even if the matrix 
coordinates are. However, by restricting the coefficients, it is possible to define Hermitian 
matrix functions, that is to say: functions whose image is a Hermitian matrix if the arguments 
are. Moreover, the sum and the composition of two Hermitian matrix functions is again a Hermitian 
matrix function. For simplicity, we will not require Hermiticity. So, from now on, we will allow 
the matrix coordinates $X^{\mu}$ to be general complex matrices. We will give more thought about 
Hermiticity in the discussion in section \ref{sect_discussion}.

\section{The substitution operator and matrix scalars}
\label{sect_scalar}

In this section we will define a matrix version of the operator $\xi^{\rho}\partial_{\rho}$.
In the Abelian case, an infinitesimal coordinate transformation looks like
\be
x^{\mu} \rightarrow y^{\mu} = x^{\mu} - \xi^{\mu}(x).\label{ct}
\ee
The parameter $\xi(x) = (\xi^0(x),...,\xi^{D-1}(x))$ is a set of $D$ Abelian functions 
$\xi^{\mu}(x)$ that transform as a vector under coordinate transformations, where $D$ is the number 
of space-time dimensions. The matrix analogy of the general coordinate transformation (\ref{ct}) 
looks like
\be
X^{\mu} \rightarrow Y^{\mu}(X) = X^{\mu} - \Xi^{\mu}(X)\,,\label{mct}
\ee
where the parameter $\Xi(X) = (\Xi^0(X),...,\Xi^{D-1}(X))$ consists of $D$ matrix functions 
$\Xi^{\mu}(X)$\,. Notice that we do not (yet) call $\Xi$ a vector, as being a vector implies a 
certain transformation under matrix coordinate transformations. At this point, we do not know 
yet how a matrix generalization of the vector transformation looks like.

An Abelian scalar is defined as a function of the coordinates which is invariant under coordinate 
transformations~(\ref{ct}):
\be
f'(y) = f(x).
\ee
Its variation is again a scalar and defined by
\be
\delta_{\xi} f(x) = f'(x) - f(x)\,.\label{a_scalar}
\ee
To first order in $\xi$, the variation is equal to
\bea
\delta_{\xi} f(x) \ = \ \sum_{k=0}^{\infty}\frac{1}{k!}
   \Bigl(\partial_{\lambda_1}...\partial_{\lambda_k}f \Bigr)(0)
           \Bigl(\xi^{\lambda_1}x^{\lambda_2}...x^{\lambda_k} + ... 
            + x^{\lambda_1}...x^{\lambda_{k-1}}\xi^{\lambda_k}\Bigr)
\ = \ \xi^{\rho}\partial_{\rho}f(x)\,\label{a_expansion}.
\eea
We will now use the same reasoning to construct a matrix scalar and its variation: a matrix 
scalar is a matrix function which is invariant under MCTs:
\be
F'(Y) = F(X).
\ee
The variation of the matrix scalar is defined by $F'(X) - F(X)$ and to first order in $\Xi^{\mu}$, 
it can be written in function of the series expansion as
\be
\delta_{\Xi}F(X) = \sum_{k=0}^{\infty}\frac{1}{k!}f_{\lambda_1...\lambda_k}
\Bigl(\Xi^{\lambda_1}(X) X^{\lambda_2}...X^{\lambda_k} \ + \ ... 
       \ + \ X^{\lambda_1}...X^{\lambda_{k-1}}\Xi^{\lambda_k}(X)\Bigr)\,.
\ee
We can read this expansion as being the expansion of $F(X)$, where each $X$ in turn is substituted 
by a $\Xi(X)$ in the same place, and summed over all possibilities, taking in account
that each $\Xi^\lambda(X)$ is again a matrix function of the type (\ref{m_function}). Actually, 
this is the same 
procedure as in the Abelian case shown by the expansion in (\ref{a_expansion}), only that there 
the procedure is equivalent to taking a derivative $\partial_\rho$ followed by a multiplication 
by $\xi^{\rho}$. This equivalence, however, is not true in the matrix case, due to the non-Abelian 
character of the matrix coordinates. Instead, we define an operator 
$\overline{\Xi^{\rho}\partial_{\rho}}$ 
which is read as follows: take each $X^{\rho}$ in turn and substitute by a $\Xi^{\rho}(X)$. The 
upper line indicates the non-Abelian nature of this substitution operator. 
The effect of the substitution operator on a matrix 
scalar is determined by the expansion:
\be
\overline{\Xi^{\rho}\partial_{\rho}}F(X) = \sum_{k=0}^{\infty}\frac{1}{k!}f_{\lambda_1...\lambda_k}
\Bigl(\Xi^{\lambda_1}(X) X^{\lambda_2}...X^{\lambda_k} 
          \ +\ ... \ + \ X^{\lambda_1}...X^{\lambda_{k-1}}\Xi^{\lambda_k}(X) \Bigr)
\label{m_subst}
\ee
Having defined the substitution operator, we can simply say that the variation of the matrix 
scalar is
\be
\delta_{\Xi}F(X) = \overline{\Xi^{\rho}\partial_{\rho}}F(X).\label{m_scalar_transf}
\ee
The commutator of two substitution operators is again a substitution operator,
\be
\overline{\Xi^{\rho}\partial_{\rho}}\bigl( \overline{\Lambda^{\sigma}\partial_{\sigma}}F(X)\bigr) 
- \overline{\Lambda^{\rho}\partial_{\rho}}\bigl( \overline{\Xi^{\sigma}\partial_{\sigma}}F(X)\bigr) 
= \overline{(\overline{ \Xi^{\rho}\partial_{\rho}}\Lambda^{\sigma} 
- \overline{\Lambda^{\rho}\partial_{\rho}}\Xi^{\sigma} )\partial_\sigma}F(X),
\label{m_scalarcommutator}
\ee
with parameter $\overline{ \Xi^{\rho}\partial_{\rho}}\Lambda^{\sigma} 
- \overline{\Lambda^{\rho}\partial_{\rho}}\Xi^{\sigma}$. This shows that the substitution operators 
form an algebra. Note that for the case that the embedding scalars commute, our results reduce 
to the known results of the Abelian case.

Since the substitution operator works linearly on the series expansion, the sum of two matrix 
scalars is again a matrix scalar. Moreover, the product of two matrix scalars is also a matrix 
scalar. The Leibniz rule holds in this case:
\be
\overline{\Xi^{\rho}\partial_{\rho}}(F\cdot G) = \overline{\Xi^{\rho}\partial_{\rho}}F \cdot G 
+ F \cdot \overline{\Xi^{\rho}\partial_{\rho}} G.
\ee

\section{The contravariant matrix vector}\label{sect_contra}

In the Abelian case, a contravariant vector consists of $D$ components $a^{\mu}(x)$, which in 
turn are functions of the coordinates. Under an infinitesimal coordinate transformation 
with parameter $\xi^{\rho}$, the contravariant vector transforms as
\be
\delta_{\xi} a^{\mu} = \xi^{\rho}\partial_{\rho}a^{\mu} - a^{\rho}\partial_{\rho}\xi^{\mu}.
\ee
Using the substitution operator~(\ref{m_subst}), the definition of a contravariant vector can 
easily be adapted to the matrix case. A matrix vector consists of $D$ components $A^{\mu}(X)$, 
which are matrix functions of the matrix coordinates, their expansion being given by
\be
A^\mu  = \sum_{k=0}^{\infty}\, \frac{1}{k!}\, a^\mu_{\lambda_1...\lambda_k}
                       \ X^{\lambda_1}...\,X^{\lambda_k}\,.
\label{m_vector}
\ee 
The variation of this covariant matrix 
vector is again a matrix covariant vector and given by
\be
\delta_{\Xi} A^{\mu} = \overline{\Xi^{\rho}\partial_{\rho}}A^{\mu} 
- \overline{A^{\rho}\partial_{\rho}}\Xi^{\mu}.\label{m_vector_trans}
\ee
The commutator of two variations is then 
\bea
[\delta_{\Xi}, \delta_{\Lambda}] A^{\mu} &=& \overline{\Xi^{\rho}\partial_{\rho}}
\Bigl( \overline{\Lambda^{\sigma}\partial_{\sigma}}A^{\mu} - \overline{A^{\rho}\partial_{\rho}}
\Lambda^{\mu}\Bigr) \nonumber \\
&=& \overline{\Bigl( \overline{\Xi^{\rho}\partial_{\rho}}\Lambda^{\sigma} 
- \overline{\Lambda^{\rho}\partial_{\rho}}\Xi^{\sigma}\Bigr)\partial_{\sigma}}A^{\mu} 
- \overline{A^{\rho}\partial_{\rho}}\Bigl( \overline{\Xi^{\sigma}\partial_{\sigma}}\Lambda^{\mu} 
- \overline{\Lambda^{\sigma}\partial_{\sigma}}\Xi^{\mu} \Bigr),
\eea
which is again of the same form as (\ref{m_vector_trans}), with parameter 
$\overline{\Xi^{\rho}\partial_{\rho}} \Lambda^{\sigma} 
- \overline{\Lambda^{\rho}\partial_{\rho}}\Xi^{\sigma}$, indicating 
that the matrix vector transforms indeed under a vector-like representation of the MCTs.

Moreover, the matrix vector transformation is linked to the matrix scalar transformation by the 
commutator. Indeed,
\bea
\delta_{\Xi}( \delta_{\Lambda}\, F ) &=& \overline{\Bigl( \overline{\Xi^{\rho}\,\partial_{\rho}}\,
\Lambda^{\sigma} - \overline{\Lambda^{\rho}\,\partial_{\rho}}\,\Xi^{\sigma} \Bigr) \partial_{\sigma}}F 
+ \delta_{\Lambda}(\delta_{\Xi}\,F)\nn
&=& \delta\,_{(\delta_{\Xi}\Lambda^\mu)}\,F + \delta_{\Lambda}(\delta_{\Xi}\,F)\,.\label{m_scalar_vector}
\eea
This relation is the matrix version of the relation between Abelian scalar and vector 
transformations
\be
\delta_{\xi}(\delta_{\lambda}\,f) = \delta\,_{(\delta_{\xi}\lambda^\mu) }\,f 
                               + \delta_{\lambda}(\delta_{\xi}\,f)
\,.\label{a_scalar_vector}
\ee

While the definition and the transformation of the matrix contravariant vector seem promising, 
some properties of Abelian vectors are not met by the matrix vector. Indeed, one of the most 
basic properties of an Abelian vector is that it belongs to a vector space. While this property 
may seem trivial, it can not be met by a matrix vector (at least not over the matrix scalars), 
regardless how the latter is defined. Due to their non-commutative nature, the matrix scalars 
do not form a field. Matrix vectors can thus at most form a module over the matrix scalars, but 
not a vector space.
Suppose we left multiply the matrix vector $A^{\mu}$ by the matrix scalar $F$. Assuming the 
Leibniz rule holds, the variation of such a product is:
\bea
\delta_{\Xi}(F\cdot A^{\mu}) &=& \delta_{\Xi}F \cdot A^{\mu} + F \cdot \delta_{\Xi}A^{\mu}\nonumber\\
&=& \overline{\Xi^{\rho}\partial_{\rho}} (F \cdot A) - F \cdot \overline{A^{\rho}\partial_{\rho}}\Xi.
\eea
The variation of the product is clearly different from the matrix vector transformation shown in 
eq. (\ref{m_vector_trans}). This may mean that we have to abandon the Leibniz rule and define the 
variation of $F \cdot A^{\mu}$ to be of the form described in eq.~(\ref{m_vector_trans}). There 
are, however, two other possibilities, both holding the Leibniz rule. The first is stating that 
the scalar multiplication of a matrix vector does not necessarily result in a matrix vector. The 
second possibility is to define a bigger set of matrix vectors, closed under both left and right 
scalar multiplication. These objects are linear combinations of objects like
 $F \cdot A^{\mu}\cdot G$. The transformation of such objects is not straightforward and should be 
defined by the Leibniz rule. It is still unclear whether these new objects should be included in 
the definition of matrix vectors or not. For the sake of clarity, we will not do so. Whenever 
we refer to a matrix contravariant vector, it will be an object of the form $A^{\mu}$.

\section{Covariant vectors and tensors}\label{sect_co}

While the contravariant vectors are linked to the scalars by the commutator, the covariant 
vectors are linked by the differential operator.
In the Abelian case,
\be
df = dx^{\rho}\partial_{\rho}f.
\ee
The differential operator $d = dx^{\rho}\partial_{\rho}$ can easily be extended to the matrix 
case using the substitution operator. We define the matrix operator $\ud$ acting on a scalar
function as converting in turn each $X$ in the expansion into a $\ud X$ and summing over all 
contributions:
\bea
\ud F \ =\ \sum_{k=0}^{\infty}\,\frac{1}{k!}\,f_{\lambda_1...\lambda_k}\, 
\Bigl(\ud X^{\lambda_1} X^{\lambda_2}... X^{\lambda_k} + ... 
           + X^{\lambda_1}...X^{\lambda_{k-1}}\ud X^{\lambda_k}\Bigr)
\ =\ \overline{\ud X^{\rho}\,\partial_{\rho}}\,F\,.
\eea
Here, $\ud X^{\mu}$ is seen as a basis element for a sort of matrix cotangent space. We assume 
that $\ud X^{\mu}$ does not commute with $X^{\nu}$. As in the Abelian case, however, we assume
that the operator $\ud$ commutes with $\delta$. The object $\ud F$ transforms then as follows:
\bea
\delta_{\Xi}\,(\ud F) 
     \ =\ \ud\, (\delta_{\Xi}F)
     \ = \  \ud\, (\overline{\Xi^{\mu}\,\partial_{\mu}}\,F)
     \ = \ \overline{\ud \Xi^{\mu}\,\partial_{\mu}}\,F + \overline{\Xi^{\mu}\,\partial_{\mu}}\,\ud F\,,
\label{udF}
\eea
where it is understood that in the last term $\overline{\Xi^{\mu}\partial_{\mu}}$ only acts on 
the $X$'s in the expansion of $\ud F$, and leaves the $\ud X$'s untouched. In order to interpret
(\ref{udF}) as a variation of $\ud F$ without reference to $F$ itself, we should be able to write
the right hand side in function of  $\ud F$ only. 

For this purpuse, let us define $\overline{\ud \Xi^{\mu}\upartial_\mu}$ (with underlined $\partial$)
as an operator which takes in turn each $\ud X^{\mu}$ and substitutes it by $\ud \Xi^{\mu}$, leaving 
all $X$'s untouched. The operator $\overline{\ud \Xi^{\mu}\partial_{\mu}}$ can then be written as a 
composition of the differential operator $\ud$ followed by $\overline{\ud \Xi^{\mu}\upartial_\mu}$ 
and hence allows us to write the transformation~(\ref{udF}) in function of $\ud F$ only:
\be
\delta_{\Xi}(\ud F) = \overline{\ud \Xi^{\mu}\upartial_{\mu}}(\ud F) 
+ \overline{\Xi^{\mu}\partial_{\mu}}(\ud F).
\ee
With this result we can define now a general covariant matrix vector (not necessarily the 
differential of a matrix scalar) by the series
\bea
B &=& \sum_{k=1}^{\infty}\sum_{j=1}^{k}b^{(j)}\,_{\lambda_1...\lambda_k}X^{\lambda_1}... X^{\lambda_{j-1}}
\ud X^{\lambda_j}X^{\lambda_{j+1}}...X^{\lambda_{k}}\,,
\label{m_covector_series},
\eea
where the upper latin index in the expansion coefficients indicates the position of the $\ud X$ 
in the terms of the expansion. In analogy of (\ref{udF}), the variation of $B$ under coordinate 
transformations is defined as 
\be
\delta_{\Xi}\,B = \overline{\ud \Xi^{\mu}\,\upartial_{\mu}}\,B + \overline{\Xi^{\mu}\,\partial_{\mu}}
\,B\,.\label{m_vector_transf}
\ee
Taking a commutator of two such transformations with parameters $\Xi^{\mu}$ and $\Lambda^{\mu}$ 
yields a new transformation with parameter $\overline{\Xi^{\rho}\partial_{\rho}}\Lambda^{\mu} 
- \overline{\Lambda^{\rho}\partial_{\rho}}\Xi^{\mu}$.

Notice that, unlike the contravariant vector of section \ref{sect_contra}, the covariant matrix 
vector $B$ can not be written in function of $D$ components. On the other hand, when we left or 
right multiply $B$ by a matrix scalar, the product will still be of the form 
(\ref{m_covector_series}). The covariant matrix vectors form a left and right module over the 
matrix scalars.

Covariant tensors can be defined easily in the same way as the covariant vector. We have to take 
into account one subtlety. In the Abelian case, a covariant two-tensor has the form
\be
C = C_{\mu\nu}\,dx^{\mu}\otimes dx^{\nu},
\ee
The $dx^{\mu}$ is placed first to indicate that it is a basis element of the first component of 
the cotangent space. In the matrix case, the place of $\ud X^{\mu}$ has already a meaning due to 
the noncommutative nature of the factors. So, we will use a subindex $\ud_{(i)}X^{\mu}$ to indicate 
that it is a basis element of the $i$th factor of the tensor space. A general matrix covariant 
2-tensor can be defined by a series
\bea
C &=& \sum_{k=2}^{\infty}\ \sum_{\substack{i,j=1 \\ i\neq j}}^k  \, c^{(i,j)}_{\mu_1...\mu_k}\, 
X^{\mu_1}...\,\ud_{(1)} X^{\mu_i}...\,\ud_{(2)} X^{\mu_j}...\, X^{\mu_k}\, .
\label{m_tensor}
\eea
Generalisation to higher rank tensor is straightforward.

The transformation of such a tensor is given by
\be
\delta_{\Xi}C = \overline{\Xi^{\rho}\,\partial_{\rho}}\,C + \overline{\ud\Xi^{\rho}\,
\upartial_{\rho}}\,C\,.
\ee
Notice that this formula is the same as eq.~(\ref{m_vector_transf}). Indeed, the operator 
$\overline{\ud\Xi^{\rho}\,\upartial_{\rho}}$ takes care of both the $dX^{\mu}$s. In the Abelian 
limit, it reduces to the two terms of the tensorial transformation.

A special subset of the Abelian tensors is formed by the antisymmetric differential forms. 
Since we have a matrix version of a differential operator, we can construct a similar structure 
within the matrix tensors. First, we can define a antisymmetric form by imposing an antisymmetry
conditions on the coefficients in the expansion of (\ref{m_tensor})
\be
c^{(i,j)}_{\mu_1...\mu_k}\ = \ - c^{(j,i)}_{\mu_1...\mu_k}\, ,
\ee
such that the expansion is given by
\bea
C = \sum_{k=2}^{\infty} \sum_{\substack{i,j=1 \\ i\neq j}}^k  c^{(i,j)}_{\mu_1...\mu_k}\, 
\Bigl( X^{\mu_1}...\,\ud_{(1)} X^{\mu_i}...\,\ud_{(2)} X^{\mu_j}...\, X^{\mu_k}
\ - \  X^{\mu_1}...\,\ud_{(2)} X^{\mu_i}...\,\ud_{(1)} X^{\mu_j}...\, X^{\mu_k}\Bigr).
\label{m_twoform}
\eea
The second step is extending the differential operator $\ud$ to vectors and antisymmetric forms. 
Let us give an example with the covariant matrix vector $B$ defined in (\ref{m_covector_series}).
\bea
\ud\, B &\equiv& \sum_{k=2}^{\infty} \Bigl[ \ 
b^{(1)}_{\mu_1...\mu_k}\Bigl(\, 
                   \ud_{(1)} X^{\mu_1}\,\ud_{(2)}X^{\mu_2}\,X^{\mu_3}...\,X^{\mu_k}
               \ + \ \ud_{(1)}X^{\mu_1}\,X^{\mu_2}\,\ud_{(2)}X^{\mu_3}...\,X^{\mu_k} \nn
& & \hsp{4.3cm}
               \ +\ ...
               \ +\ \ud_{(1)}X^{\mu_1}\,X^{\mu_2}...\,X^{\mu_{k-1}}\,\ud_{(2)} X^{\mu_k}\Bigr)\nn
& & \hsp{.5cm} -\, b^{(1)}_{\mu_1...\mu_k}\Bigl(\, 
                   \ud_{(2)} X^{\mu_1}\,\ud_{(1)}X^{\mu_2}\,X^{\mu_3}...\,X^{\mu_k}  
               \ + \ \ud_{(2)}X^{\mu_1}\,X^{\mu_2}\,\ud_{(1)}X^{\mu_3}...\,X^{\mu_k} \nn
& & \hsp{4.3cm}
               \ + \ ... 
               \ +\ \ud_{(2)}X^{\mu_1}\,X^{\mu_2}...\,X^{\mu_{k-1}}\,\ud_{(1)} X^{\mu_k}\Bigr)\nn
& & \hsp{.5cm}  +  \ ...\ + \\
& &  \hsp{.5cm} + \,b^{(k)}_{\mu_1...\mu_k}\Bigl(\, 
                    \ud_{(2)} X^{\mu_1}\,X^{\mu_2}...\,X^{\mu_{k-1}}\,\ud_{(1)}X^{\mu_k}
               \ + \ X^{\mu_1}\,\ud_{(2)}X^{\mu_2}\,X^{\mu_3}...\,X^{\mu_{k-1}}\,\ud_{(1)}X^{\mu_k} \nn
& & \hsp{4.3cm} 
               \ + \ ...  
               \ + \  X^{\mu_1}...\,X^{\mu_{k-2}}\,\ud_{(2)}X^{\mu_{k-1}}\,\ud_{(1)} X^{\mu_k}\Bigr)\nn
& & \hsp{.5cm} -\, b^{(k)}_{\mu_1...\mu_k}\Bigl(\, 
                    \ud_{(1)} X^{\mu_1}\,X^{\mu_2}...\,X^{\mu_{k-1}}\,\ud_{(2)}X^{\mu_k}
               \ + \ X^{\mu_1}\,\ud_{(1)}X^{\mu_2}\,X^{\mu_3}...\,X^{\mu_{k-1}}\,\ud_{(2)}X^{\mu_k} \nn
& & \hsp{4.3cm}
               \ + \ ... 
               \ + \ X^{\mu_1}...\,X^{\mu_{k-2}}\,\ud_{(1)}X^{\mu_{k-1}}\,\ud_{(2)} X^{\mu_k}\Bigr)\Bigr]
\,.\nonumber
\eea
This definition ensures that $\ud\, (\ud\ .) = 0$. Indeed, if $B = \ud F$, then all coefficients
$b^{(j)}_{\lambda_1...\lambda_k}$ in (\ref{m_covector_series}) are equal for a given index structure
$\{\lambda_1...\lambda_k\}$, and the antisymmetry of the $\ud$ operator makes that the different
terms in $\ud B$ cancel. Also the other way around is correct: if $\ud B=0$, then this is due to the
fact that the different terms in $B$ have the same coefficients and hence $B$ can be written as
$B= \ud F$. This property is clearly important as one intents to incorporate gauge fields in the 
multiple D-brane effective action.

\section{The contravariant vector again}\label{sect_contra_2}

We have what appears to be a consistent definition for covariant vectors, tensors and form fields. 
We have also made some attempts to construct contravariant vectors. In the Abelian case, the 
covariant vectors are defined as the dual space to the contravariant vectors. Can we, at this point, define a sort of contraction between covariant and contravariant matrix vectors, resulting in a matrix scalar?

The answer lies in the substitution operator $\overline{\ud \Xi^{\rho}\,\upartial_{\rho}}$ we 
defined earlier. It substitutes $\ud X^{\rho}$ by $\ud \Xi^{\rho}$, turning a matrix vector into 
a matrix vector. If we define an operator which substitutes $\ud X^{\rho}$ by some matrix function 
instead, the result will be a matrix function without $\ud X^{\rho}$. In short, the contraction 
between
\bea
B &=& \sum_{k=1}^{\infty}\sum_{j=1}^{k}b^{(j)}_{\lambda_1...\lambda_k}X^{\lambda_1}... 
X^{\lambda_{j-1}}\ud X^{\lambda_j}X^{\lambda_{j+1}}...X^{\lambda_{k}},
\eea
and the contravariant vector $A = \overline{A^{\mu}\upartial_{\mu}}$ is
\bea
A\cdot B \ = \ \overline{A^{\mu}\upartial_{\mu}}\,B 
      \ = \ \sum_{k=1}^{\infty}\sum_{j=1}^{k}b^{(j)}\,_{\lambda_1...\lambda_k}X^{\lambda_1}... X^{\lambda_{j-1}}
              A^{\lambda_j}X^{\lambda_{j+1}}...X^{\lambda_{k}}\,,
\label{m_contraction}
\eea
where $A^\mu$ is of the form of (\ref{m_vector}). In order to check whether $A \cdot B$ behaves as a matrix scalar, one needs to calculate its variation by means of the Leibniz rule. Indeed, we see that 
\bea
\delta_{\Xi}(A\cdot B) &=& \delta_{\Xi}A\cdot B + A \cdot \delta_{\Xi} B\nn
&=& \overline{(\overline{\Xi^{\rho}\,\partial_{\rho}}\,A^{\mu} - \overline{A^{\rho}\,\partial_{\rho}}
\,\Xi^{\mu} )\,\upartial_{\mu}}\,B + \overline{A^{\mu}\,\upartial_{\mu}}\,(\overline{\Xi^{\rho}\,
\partial_{\rho}}\,B + \overline{\ud \Xi^{\rho}\,\partial_{\rho}}\,B)\nn
&=& \overline{\Xi^{\rho}\,\partial_{\rho}}\,(A \cdot B)\,.
\eea
This proves that the covariant and contravariant matrix vectors are indeed dual objects. In fact, 
the matrix scalar transformation~(\ref{m_scalar_transf}) can be written in terms of a differential 
and a contraction with a matrix contravariant vector:
\be
\overline{\Xi^{\rho}\,\partial_\rho}\,F = \overline{\Xi^{\rho}\,\upartial_{\rho}}\,\ud F\,.
\ee
This brings the transformation in closer connection to the Abelian case, where 
$\xi^{\rho}\partial_{\rho}$ can be interpreted as a derivative followed by a contraction.

Every form dual to $\overline{A^{\rho}\,\upartial_{\rho}}$ is of the form defined by formula 
(\ref{m_covector_series}). On the other hand, there are objects dual to the matrix covariant 
vector which are not of the form $\overline{A^{\rho}\upartial_{\rho}}$. Indeed, take any operator 
$F \cdot \overline{A^{\mu}\upartial_{\mu}} \cdot F'$ which works on a matrix covariant vector $B$ 
as follows. Take every $dX^{\mu}$ out of the series of $B$ and replace it with a $A^{\mu}$, then 
left multiply the result by a matrix scalar $F$, then right multiply the result by a matrix 
scalar $F'$:
\be
(F\cdot \overline{A^{\rho}\upartial_{\rho}}\cdot F')\,B = F \cdot (\overline{A^{\mu}\,\
\upartial_{\mu}}\,B)\cdot F'\,.
\ee
The result of this operation is a matrix scalar, as it is the product of three matrix scalars.
So, the dual space of the matrix covariant vectors consists not only of matrix contravariant vectors of the form 
$\overline{A^{\mu}\upartial_{\mu}}$, but there are also linear combinations of operators such as 
$F\cdot \overline{A^{\mu}\,\upartial_{\mu}}\cdot F'$, which, as mentioned in section 
\ref{sect_contra}, do not behave as contravariant vector in the sense of the definition used there.
Indeed, the set of these operators is closed under left and right multiplication by matrix scalars. 
As a drawback, the operations described here have a difficult variation.

Now we are able to spot a more serious problem, namely trying to identify the dual space of this 
second, bigger set of matrix contravariant vectors. Its elements would be linear combinations of 
operators like $G \cdot B \cdot G'$, working on $F\cdot \overline{A^{\mu}\upartial_{\mu}}\cdot F'$ 
like
\be
(G \cdot B \cdot G')(F\cdot \overline{A^{\mu}\upartial_{\mu}}\cdot F') = G \cdot F \cdot 
\overline{A^{\mu}\upartial_{\mu}}B \cdot F' \cdot G'.
\ee
From these examples it becomes clear what we are getting into: a tower of ever bigger spaces for 
the matrix vectors. As things are now, no solution has yet been found to this problem.
The same difficulty can also be spotted in a different way. If we compare the set of the matrix 
contravariant vectors $A^{\rho}\,\upartial_{\rho}$ to the set of the matrix covariant vectors 
defined by the series~(\ref{m_covector_series}), we see that they are not isomorphous. This 
problem needs to be solved before we can define a matrix generalization of a metric, which 
implies an isomorphism between covariant and contravariant vectors.

\section{Discussion}\label{sect_discussion}

At first sight, the substitution operator seems to be a valid candidate for infinitesimal MCTs. 
Indeed, the substitution operators do form an algebra. A consistent definition of a matrix scalar 
has been given in section \ref{sect_scalar}. The product of two matrix scalars is a matrix scalar 
and its variation obeys the Leibniz rule.

A matrix generalization of the contravariant vector has been defined in section \ref{sect_contra}. 
Its variation~(\ref{m_vector_trans}) has been modelled after the variation of the Abelian 
contravariant vector. With this definition, the transformation of the contravariant matrix
 vector is linked to the transformation of the matrix scalar by the commutation relation given 
in eq.~(\ref{m_scalar_vector}). Moreover, the parameter of the infinitesimal MCT varies in the 
same way as the matrix vector. Unlike the Abelian vectors, however, the matrix vectors do not 
form a vector space. It is actually impossible to define matrix vectors such that they would 
form a vector space over the matrix scalars, as the matrix scalars do not form a field. If we 
can not make a vector space, can we at least form a module? The answer is quite tricky, as 
there are different possibilities. We can make the set of the contravariant matrix vectors 
defined by $A^{\mu}$ and with a transformation given by eq.~(\ref{m_vector_trans}) into a 
module if we abandon the Leibniz rule. Losing the Leibniz rule is undesirable, especially since 
the rule holds for the scalars. The preferred idea is that scalar multiplication of a 
contravariant matrix vector results in a new object, which is more complex than the original 
contravariant matrix vector. The variations of these objects are determined by the Leibniz rule and 
are quite cumbersome. It is still unclear whether these new objects should be included in the 
definition of matrix vectors or not.

We have defined matrix covariant vectors in section \ref{sect_co} using a generalization of 
the differential operator. First, we have constructed a matrix generalization of a differential 
operator $\ud$ similar to the substitution operators defined earlier. The action of this $\ud$ 
on a matrix scalar function is based on the action of the De Rham differential operator on an 
Abelian scalar. Assuming that the variation of a differential is equal to the differential of 
the variation, the variation of $\ud F$ turns out to be of the form~(\ref{udF}). We defined a 
series expansion and a variation for a general covariant matrix vector on basis of this $\ud F$. 
It turns out that the contraction of a contravariant matrix vector $A^{\mu}$ with a covariant 
matrix vector is a matrix scalar indeed. The Leibniz rule holds in this case, which strengthens 
our argument against abandoning it for scalar multiplication of matrix contravariant vectors. 
On top of this, the matrix covariant vectors do form a left and a right module under 
multiplication with matrix scalars. 

It is possible to extend the definition of the matrix differential operator $\ud$ towards 
matrix covariant vectors, resulting in matrix two-tensors. By calculating $\ud^2$, we prove 
that $\ud$ indeed behaves as an exterior derivative. Higher-rank matrix tensors and antisymmetric 
forms were constructed in section \ref{sect_co}.

There seem to be quite some differences between the matrix covariant and contravariant vectors. 
Unlike the matrix contravariant vector, the matrix covariant vectors can not be written in terms 
of $D$ components which are matrix functions. The matrix covariant vectors form a module for 
scalar multiplication while the matrix contravariant vectors do not. It is impossible to define 
an isomorphism between the matrix contravariant and covariant vectors. Indeed, the two sets 
have a different cardinality. This means we can not define a matrix generalization of a metric. 
It is possible that there exist better definitions for the matrix vectors than ours, which do 
allow a matrix generalization of a metric.
The same problem has been discussed from a different approach in section \ref{sect_contra_2}. 
The covariant vectors defined by the series in  (\ref{m_covector_series}) form the entire dual 
space to the contravariant vectors. The opposite, however, is not true, indicating also the difference 
in cardinality.

Just like Abelian coordinates are real variables, the matrix coordinates are Hermitian matrices. 
In the previous sections, we allowed the matrix coordinates to be general complex matrices. It 
is, however, quite easy to restrict the entire construction to Hermitian matrix coordinates. By 
restricting the coefficients in the series (\ref{m_function}), we can define a Hermitian matrix 
function. Such a function takes values in the Hermitian matrices if its arguments are Hermitian. 
If the parameter of an infinitesimal MCT consists of $D$ Hermitean matrix functions, it 
transforms a Hermitian matrix coordinate into a Hermitian matrix coordinate. Moreover, a 
substitution with such a parameter will also conserve the Hermiticity of matrix functions. 
This allows a restriction of the discussion to Hermitian matrix coordinates and objects.

Finally we want to indicate how the substitution operators can be useful in combination with 
an inclusion function $\Phi$. This inclusion function is meant to map the Abelian functions 
onto the matrix functions. As it is a homomorphism, it will transport all structures from the 
Abelian case, including commutativity. In this case, we expect no problems with the corresponding 
contravariant and covariant matrix vectors in the image of the inclusion function. The image 
of the inclusion function is a subset of the matrix functions defined in (\ref{m_function}). 
The coefficients $f_{\lambda_1...\lambda_k}$ are determined by the expansion coefficients of the 
corresponding Abelian function and the background metric. As a side remark, the inclusion 
function is linked to an Abelian background while the original set-up in our article is not. 
The matrix transformations need to be modified to work on the background metric as well as the 
matrix coordinates. Indeed, suppose that
\be
F(X,g) = \Phi(f)(X,g)
\ee
is the image of the Abelian scalar $f(x)$. Then the variation of $F(X,g)$ under a MCT with 
parameter $\Xi^{\rho}(X,g) = \Phi(\xi^{\rho})(X,g)$ is given by
\be
(\delta_{\Xi}F)(X,g) = \overline{\Xi^{\rho}\partial_{\rho}}F(X,g) 
- \Subst(g \rightarrow \delta_{\xi}g)F(X,g).
\ee
Here, $\Subst(g \rightarrow \delta_{\xi} g)$ means a substitution of the background metric in 
the expansion by $\delta_{\xi}g$, in the same way as $X^{\mu}$ is substituted by $\Xi^{\mu}$ under 
the action of $\overline{\Xi^{\mu}\partial_{\mu}}$.


\end{document}